\documentstyle[preprint,aps]{revtex}
\tightenlines
\begin{document}
\newcommand{\kbar}{\mbox{$k\hspace{-.09in}{}^-$}}
\draft
\title{Quantum and classical chaos for a single trapped ion}
\author{A.J.Scott~$^\dagger$, C.A.Holmes~$^\dagger$ and G.J.Milburn~$^*$}
\address{$^\dagger$~Department of Mathematics,\\
$^*$~Department of Physics,\\
The University of Queensland,
QLD 4072 Australia.}
\maketitle

\begin{abstract}
 In this paper we investigate the
quantum and classical dynamics of a single trapped ion subject to
non-linear kicks derived from a periodic sequence of
Gaussian laser pulses.  We show that the classical system exhibits
diffusive growth in the energy, or 'heating', while quantum
mechanics suppresses this heating.  This system may be realised in current
single trapped-ion experiments with the addition of
near-field optics to introduce tightly focussed laser pulses into the trap.

\end{abstract}

\newpage
\section{Introduction}
Recent experiments on the nonlinear dynamics of cold trapped atoms have
provided a remarkable verification of key theoretical
ideas in the subject of quantum chaos\cite{Raizen95,Christensen98},
including dynamical localisation and the effect of
decoherence in restabilising the classical limit. In all these experiments
however the observed results are obtained from a
large ensemble of single atom experiments which run in parallel but
independently.  Until now, there have been no experiments
which investigate quantum chaos of a {\em single} particle monitored over a
period of time. In contrast, in the related field
of ion-trapping, technological advances now enable a single ion to be
trapped, cooled to the ground state of the trap, and
monitored, almost without error\cite{NIST}. The quantum dynamics of the
centre of mass motion of the ion is extremely well
described by a three dimensional harmonic potential. In some experiments,
two degrees of freedom are very tightly bound and
the interesting harmonic motion takes place in a single degree of freedom.
Of course this system is integrable. However if
this degree of freedom is subject to  a periodic nonlinear potential, chaos
may result. In this paper we investigate the
quantum and classical dynamics of a single trapped ion subject to
non-linear kicks derived from a periodic sequence of
Gaussian laser pulses.   This system may be realised in current single
trapped-ion experiments with the addition of
near-field optics to introduce tightly focussed laser pulses into the trap.
Another suggestion for investigating quantum chaos in a single trapped ion has recently been suggested by
Berman et al \cite{Berman1999}.

The recently achieved ability to engineer dynamics for a single trapped ion has followed from the potential application of this system for quantum computational gates. As such these systems necessarily operate at the quantum level and provide the ideal experimental context to test quantum nonlinear dynamics. Indeed such experiments will ultimately involve the ability to follow the quantum dynamics of many trapped ions with complex many-body interactions introduced by externally imposed time dependant Hamiltonians. We thus believe it timely to consider tests of quantum chaos which can be made with current technology.

In section II we define the classical dynamical system and give a
detailed analysis of the classical motion and the
transition to chaos. In section III we give a quantum description of the
problem and  in section IV show that this system
exhibits  a suppression in the diffusion of momentum and position. In other
words the total energy of the
trapped ion is localised in contrast to classical diffusive heating of the
motion. Finally in Section V we discuss a possible
physical realisation of the system.

\section{Classical Map}
The system has the Hamiltonian
\begin{displaymath}
\tilde{H}=\frac{\tilde{p}^2}{2m}+\frac{m\Omega^2}{2}\tilde{x}^2+\kappa
e^{-\alpha \tilde{x}^2}T\sum_{n=-\infty}^{\infty}\delta(\tilde{t}-nT),
\end{displaymath}
where $m$ is the mass of the ion trapped in a harmonic potential of frequency $\Omega$. The ion is subject to a periodic sequence of laser pulses with period $T$. The $\kappa
e^{-\alpha \tilde{x}^2}$ term in the Hamiltonian describes the potential felt by the ion due to the Gaussian structure of the laser. If we rescale time, position and momentum by letting $\tilde{t}=tT$,
$\tilde{x}=\frac{1}{\sqrt{\alpha}}x$ and
$\tilde{p}=\frac{m\Omega}{\sqrt{\alpha}}p$ the Hamiltonian rescales to
\begin{equation}
H(x,p,t)=\frac{\alpha
T}{m\Omega}\tilde{H}(\tilde{x},\tilde{p},\tilde{t})=\frac{\omega}{2}\left(p^2+x^
2\right)+ke^{-x^2}\sum_{n=-\infty}^{\infty}\delta(t-n).
\label{H1}
\end{equation}
where $\omega =\Omega T$ and $k=\frac{\kappa\alpha T}{m\Omega}$ are
dimensionless parameters. The new variables $x$, $p$ and $t$ together with the Hamiltonian, $H$, are also dimensionless.

Between kicks the system has the solution $(x,p)=(x_0\cos{\omega
t}+p_0\sin{\omega t},-x_0\sin{\omega t}+p_0\cos{\omega t})$. The effect of
the kick is to add a position dependent shift in the momentum of
$2kxe^{-x^2}$. Denoting the mapping by $F$ we can write it as the
composition of a kick, $K$, and a linear rotation, $W$,
\begin{displaymath}
F=W\circ K
\end{displaymath}
where,
\begin{eqnarray*}
K(x,p) & = & (x,p+2kxe^{-x^2}) \\
W(x,p) & = & (x\cos{\omega}+p\sin{\omega},-x\sin{\omega}+p\cos{\omega})
\end{eqnarray*}
Hence $F$ maps $(x,p)$ from just before a kick to one period later.

The fixed points of $F$ are at
\begin{displaymath}
(x,p)=(0,0),\left(\pm\sqrt{\ln(k\cot\frac{\omega}{2})},\mp\tan\frac{\omega}{2}\sqrt{\ln(k\cot\frac{\omega}{2})}\right).
\end{displaymath}
The origin is stable if $k\cot\frac{\omega}{2}<1$ and
$-k\tan\frac{\omega}{2}<1$. When $k\cot\frac{\omega}{2}>1$ it becomes
unstable via a pitchfork bifurcation which creates the second two fixed
points. These are stable for $k\cot\frac{\omega}{2}<\exp
(\frac{1}{2}\csc^2\frac{\omega}{2})$.
When $-k\tan\frac{\omega}{2}>1$ the origin becomes unstable via a period
doubling bifurcation and two period 2 orbits are created at
\begin{displaymath}
(x,p)=\left(\pm\sqrt{\ln(-k\tan\frac{\omega}{2})},\mp\cot\frac{\omega}{2}\sqrt{\
ln(-k\tan\frac{\omega}{2})}\right).
\end{displaymath}
These are stable for $-k\tan\frac{\omega}{2}<\exp
(\frac{1}{2}\sec^2\frac{\omega}{2})$.

One can also find two sets of period 4 orbits which exist for
$-k\tan\omega>1$. The first set is at
\begin{equation}
(x,p)=(\pm\chi,\pm\chi\cot\frac{\omega}{2}),(\pm\chi,\mp\chi\tan\frac{\omega}{2}
)\label{p4}
\end{equation}
where $\chi=\sqrt{\ln(-k\tan\omega)}$. They are stable if
$-k\tan\omega<\exp (\frac{1}{2}|\sec\omega|)$. A second set lie in between
these orbits at
\begin{displaymath}
(x,p)=(\pm\chi,\pm\chi\cot\omega),(0,\pm\chi\csc\omega).
\end{displaymath}
These are always unstable.

It is instructive to study the process of creation and destruction of these
period 4 orbits in more detail. In Fig. 1 we have drawn a bifurcation
diagram. The period 4 orbits exist for parameter values lying in the shaded
regions. The regions of lighter shade show where the first set of orbits
(\ref{p4}) are unstable. In figures 2-6 a sequence of phase space
pictures are drawn for values of $\omega$ along the line $k=2$. Fig. 2
is for $\frac{\omega}{2\pi}=0.24$ and shows typical phase space structure
for the system. As $\omega$ is increased, the arms of the star-shaped
chaotic region grow in size. This can be seen in Fig. 3 where
$\frac{\omega}{2\pi}=0.248$. At $\frac{\omega}{2\pi}=\frac{1}{4}$ (Fig.
4) these arms are infinitely long chaotic channels which divide the
phase space into four regions, creating the period 4 orbits at infinity.
Fig. 5 shows the period 4 orbits just after they are created
($\frac{\omega}{2\pi}=0.252$). They now move towards the origin as $\omega$
is increased further. On this journey the first set (\ref{p4}) shed their
stability via a pitchfork bifurcation but then regain it before destroying
at the origin. Fig. 6 shows the orbits just before destruction
($\frac{\omega}{2\pi}=0.42$).

It is simple to see how the period 4 orbits were created at infinity when
$\omega=\frac{2\pi}{4}$. The kick has little influence on orbits here and
thus $F$ reduces to simple linear rotation with period 4. In general, it
can be shown that an orbit of period $p$ is created at infinity when
\begin{displaymath}
\omega=\frac{2\pi q}{p},
\end{displaymath}
and then destroyed at the origin when
\begin{displaymath}
\cos\omega+k\sin\omega=\cos\frac{2\pi q}{p},
\end{displaymath}
where $q$ and $p$ are natural numbers with a greatest common divisor of
one. The condition for destruction is found by looking at the eigenvalues
of the tangent map at the origin and then equating this linear rotation
with the period of the orbit.

\section{Quantum Map}
To construct the quantum map we start with the rescaled Hamiltonian
(\ref{H1}) and define a dimensionless Planck's constant, $\kbar$, via the
commutation relation for position and momentum
\begin{displaymath}
\left[x,p\right]=\frac{\alpha}{m\omega}\left[\tilde{x},\tilde{p}\right]=i\frac{\alpha\hbar}{m\omega}\equiv i\kbar.
\end{displaymath}
The time evolution of an initial state, $|\psi^n\rangle$, from just before
a kick through to one period later is given by
\begin{eqnarray}
|\psi^{n+1}\rangle & = & \exp\left(\frac{-i}{\kbar}\int_0^1 H(t)dt\right)|
\psi^n\rangle \\
& = &
\exp\left(\frac{-i\omega}{2\kbar}\left(p^2+x^2\right)\right)\exp\left(\frac{-ik}
{\kbar}e^{-x^2}\right)|\psi^n\rangle \\
& \equiv & \hat{F}|\psi^n\rangle \label{qmap}
\end{eqnarray}
Hence the Floquet operator, $\hat{F}$, defines the quantum map. Now,
defining the annihilation and creation operators to be
\begin{displaymath}
a=\frac{1}{\sqrt{2\kbar}}(x+ip) \;\;\;\;\;\mbox{and}\;\;\;\;\;
a^\dag=\frac{1}{\sqrt{2\kbar}}(x-ip)
\end{displaymath}
respectively, and using the simple harmonic oscillator eigenstates
\begin{displaymath}
|n\rangle=\frac{1}{\sqrt{n!}}{a^\dag}^n|0\rangle \;\;\;\;\;\;\;\;\;\;
n=0,1,2\ldots
\end{displaymath}
as an orthogonal basis we can rewrite equation (\ref{qmap}) as
\begin{displaymath}
c_m^{n+1}=F_{mk}c_k^n
\end{displaymath}
where $c_m^{n}=\langle m|\psi^n\rangle$ and
\begin{eqnarray*}
F_{nm} & = & \langle n|\hat{F}|m\rangle \\
& = & e^{-i\omega(n+\frac{1}{2})}\langle
n|\exp\left(\frac{-ik}{\kbar}e^{-x^2}\right)|m\rangle.
\end{eqnarray*}
The last term is found by taking the exponential of the matrix with components
\begin{eqnarray*}
\lefteqn{\langle
n|\frac{-ik}{\kbar}e^{-x^2}|m\rangle=\frac{-ik}{\kbar^{3/2}\sqrt{\pi
n!m!2^{n+m}}}\int_{-\infty}^{\infty}H_n\left(\frac{x}{\sqrt{\kbar}}\right)H_m\left(\frac{x}{\sqrt{\kbar}}\right)e^{-(1+\frac{1}{\kbar})x^2}dx} \\
& = & \frac{-k(n+m-1)(n+m-3)\cdots
1}{\kbar^{3/2}\sqrt{n!m!}}\left(\frac{-\kbar}{1+\kbar}\right)^{\frac{n+m+1}{2}}{
}_2F_1\left(-n,-m;\frac{1-n-m}{2};\frac{1}{2}(1+\frac{1}{\kbar})\right)
\end{eqnarray*}
for $n+m$ even and vanishing otherwise. Here $H_n$ are Hermite polynomials
and ${}_2F_1$ is the hypergeometric function. Note that $F_{nm}=0$ whenever
$n+m$ is odd. This means that even and odd parity states do not couple
under $\hat{F}$ and thus evolve independently.

\section{Localisation}
We now show numerically the presence of dynamical
localisation\cite{localisation} in the system.
Or, more precisely, we show that classical diffusion is suppressed when the
system is evolved quantum mechanically.
For this,
we have chosen $\omega=\pi(3-\sqrt{5})$ and $k=8$.
In this parameter regime a large chaotic sea centred at the origin consumes
the phase space (see Fig. 7). The initial state was chosen to be
$|0\rangle$,
which has a Husimi probability density of
\begin{equation}
\left|\langle z|0\rangle\right|^2=e^{-|z|^2}\label{husimi}
\end{equation}
in phase space. Here $z=\frac{1}{\sqrt{2\kbar}}(x+ip)$ and $|z\rangle$ are
the coherent states defined as
\begin{displaymath}
|z\rangle=e^{-\frac{1}{2}|z|^2}\sum_{n=0}^{\infty}
\frac{z^n}{\sqrt{n!}}|n\rangle.
\end{displaymath}
Thus our initial state is a highly localised Gaussian hump centered at the
origin.
This was then evolved forward using 1800 of the even basis states. Fig. 8
shows the average dimensionless energy, $\langle x^2+p^2\rangle$, after each kick. The light gray is for $\kbar=0.5$ and the dark gray is for $\kbar=0.2$. The
energy under classical evolution is shown in
black. Here an initial density (\ref{husimi}) with $\kbar=0.2$ was chosen.
One can clearly see in this figure that diffusion
is suppressed after about 100 kicks. The procedure was repeated using only
1000 even basis states to confirm accuracy and it
was found that the difference in the energies did not exceed $10^{-8}$
until after 1000 kicks. Thus the localisation is truly
a property of quantum mechanics.

\section{Discussion and Conclusion}
What are the physical requirements to realise this system in current ion
trap experiments? Consider a $\mbox{}^9Be^+$ ion
such as used in the NIST experiments\cite{NIST}, with a harmonic frequency
in the relevant direction of $\Omega=1$ MHz. The key
parameter which determines the effective Planck constant is the parameter
$\alpha$. If one uses a focused laser beam, a typical
value for this parameter is $10^{10}\ \mbox{m}^2$ and a resulting effective
Planck's constant of
$\kbar=6.7\times 10^{-5}$, which is hopelessly too small. On the other hand
if we use a near-field probe, as used in
near-field optical scanning microscopy (NOSM), to inject the field we can get  a value as 
high as $\alpha=10^{14}\ \mbox{m}^2$ with a
typical probe tip of diameter
$10\ \mbox{nm}$. This corresponds to an effective Planck's constant for
$\mbox{}^9Be^+$  of
about $\kbar=0.7$, which is more promising. To achieve a value for the kick
parameter of the order used above we would need to
focus a few nanowatts into the NOSM probe which is quite typical. This
would correspond to an intensity of about $1000\
\mbox{mW cm}^{-2}$ at the ion. If we choose the kick period to be of the
order of $10\ \mbox{ms}$, the kick parameter, $k$, has a
value of the order of unity. We conclude that this experiment is possible
for a current single trapped-ion experiment with the
addition of near-field optical fibre probes.

The next question we need to ask of such a system is how are we to observe
the motion of the ion? Fortunately the current
single trapped-ion experiments are designed precisely to enable careful
monitoring of the motion states. The details are
described in reference \cite{NIST}. The  basic idea is to map the motion
states onto particular internal states of the ion
which are then probed by a fluorescent shelving technique. In particular it is 
possible to measure the centre-of-mass energy of
the ion in the trap. Each measurement however destroys the quantum state of
the  ion at that time, so repreparation
of the ion initial state is required. One then needs to perform repeated
experiments for differing number of kicks before
reading out the centre-of-mass energy. In this way it is possible to
monitor the energy of the motion as a function of kick
number. Dynamical localisation of the motion energy of the ion could thus
be observed.

 Finally we need to ask if it is feasible to prepare the initial states
we have used in this paper. Again reference
\cite{NIST} shows that it is possible to prepare the ion in the ground
state of the harmonic trap, so this part is relatively
easy. Laser pulses may then be used to displace this minimum uncertainty
state anywhere in the phase plane. This ability to
place a localised state anywhere in the phase plane would enable a
detailed study of mixed chaotic and regular phase space
structures.   Unfortunately in the
current experiment unwanted stray linear potentials cause
a heating of the ion and thus it does not stay in the ground state for
long, but rather undergoes a diffusive motion in the
phase plane\cite{Schneider99,James98}. A very considerable amount of effort
is currently being devoted to removing this
unwanted heating so that trapped ions can be used in a quantum logic gate.
We thus expect this problem to be solved or at least
significantly mitigated.

Needless to say this is not an easy experiment. Introducing the near field
probe close to the ion will cause additional
unwanted van der Walls forces to be exerted on the ion. However these
forces, while making a detailed comparison to experiment
more difficult,  will not effect the generic transition to chaos described
above so long as they remain weak. The heating of the
ion due to stray linear potentials will remain a problem to some extent.
Such fluctuating forces are a source of
decoherence and thus will tend to destroy localisation. Taking a longer
view however the ease with which decoherence can be
induced via this mechanism should enable a detailed study of the effect of
noise on dynamical localisation to be made, thus
turning a bug into a feature.

\section{Figure captions}

{\bf FIG. 1.} Bifurcation diagram for the period 4 orbits (\ref{p4}). The
stable region is in dark and the unstable in light.

\noindent
{\bf FIG. 2.} Phase portrait for $k=2$ and $\frac{\omega}{2\pi}=0.24$.

\noindent
{\bf FIG. 3.} As for Fig. 2 except $\frac{\omega}{2\pi}=0.248$.

\noindent
{\bf FIG. 4.} As for Fig. 2 except $\frac{\omega}{2\pi}=\frac{1}{4}$.

\noindent
{\bf FIG. 5.} As for Fig. 2 except $\frac{\omega}{2\pi}=0.252$.

\noindent
{\bf FIG. 6.} As for Fig. 2 except $\frac{\omega}{2\pi}=0.42$.

\noindent
{\bf FIG. 7.} Phase portrait for $k=8$ and $\omega=\pi(3-\sqrt{5})$.

\noindent
{\bf FIG. 8.} The average dimensionless energy after each kick for the initial
state $|0\rangle$ with $\kbar=0.5$ (light gray), $\kbar=0.2$ (dark gray)
and classical evolution (black).


\begin{references}

\bibitem{Raizen95}F.L.~Moore, J.C.~Robinson, C.F.~Bharucha, Bala Sundaram,
and M.G.~Raizen,Phys. Rev. Letts. {\bf 75},
4598, (1995) .

\bibitem{Christensen98} H.~Ammann, R.~Gray, I.~Shvarchuck, and
N.~Christensen, Phys. Rev. Letts. {\bf 80}, 4111 (1998).


\bibitem{NIST}D.J.~Wineland, C.~Monroe, W.M.~Itano, D.~Leibfried,
B.E.~King, and D.M.~Meekhof, "Experimental issues
in coherent quantum-state manipulation of trapped atomic ions",  Journal of
Research of the National
Institute of Standards and Technology {\bf 103},259 (1998).

\bibitem{Berman1999}G.P.~Berman, D.F.V.~James, R.J.~Hughes, M.S.~Gulley, M.H.~Holzscheiter, G.V.~Lopez, LANL eprint, quant-ph/9903063.


\bibitem{localisation}S.Fishman, D.R.Grempel and R.E.Prange, Phys. Rev. A
{\bf 29}, 1639 (1984);G.~Casati, B.V.~Chirokov,
D.L.~Shepelyansky and I.~Guarneri, Phys. Rep. {\bf 154}, 77 (1987).

\bibitem{Schneider99}S.~Schneider and G.J.~Milburn, Phys. Rev. A {\bf 59}, 3766 (1999).

\bibitem{James98}D.F.V.~James, Phys. Rev. Lett. {\bf 81} 317 (1998).

\end{references}
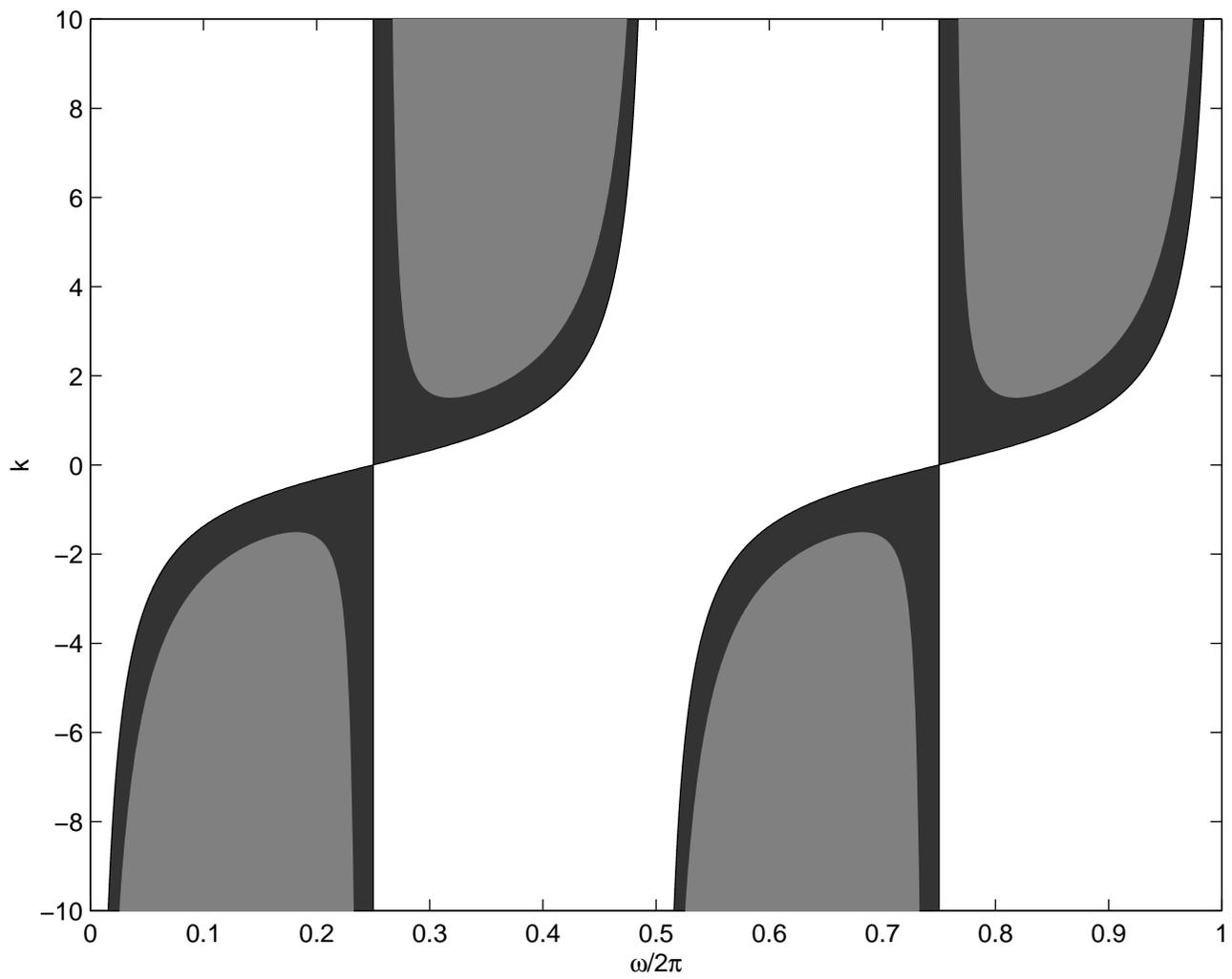
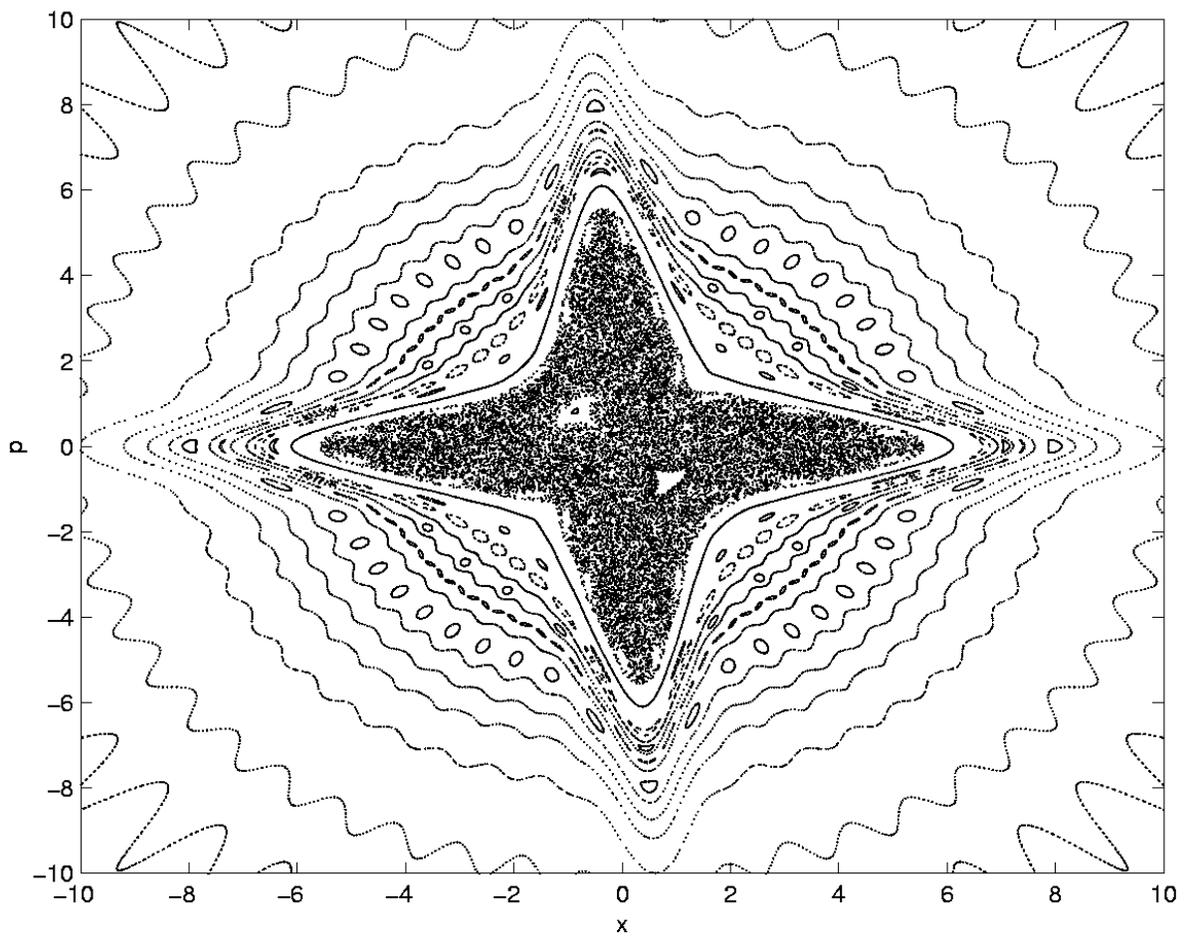
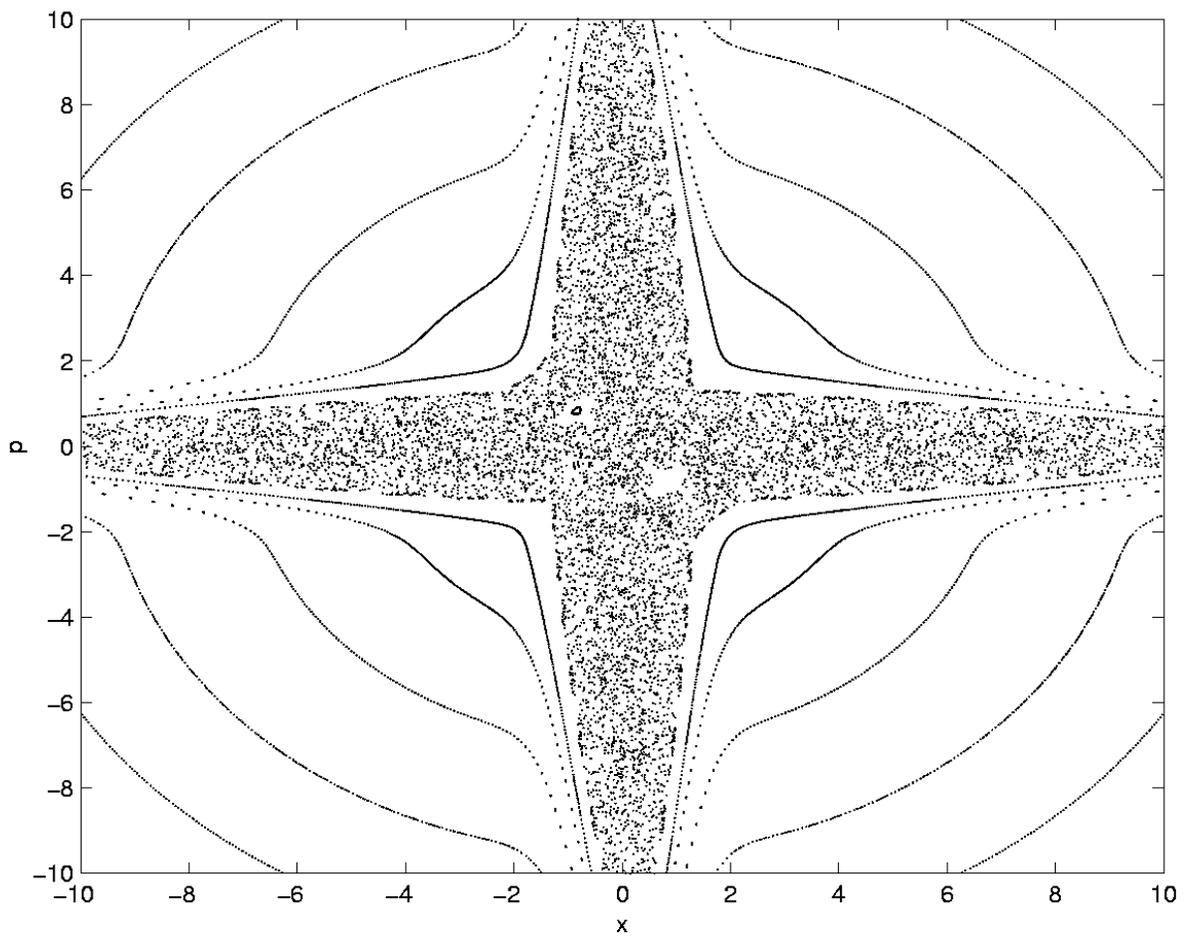
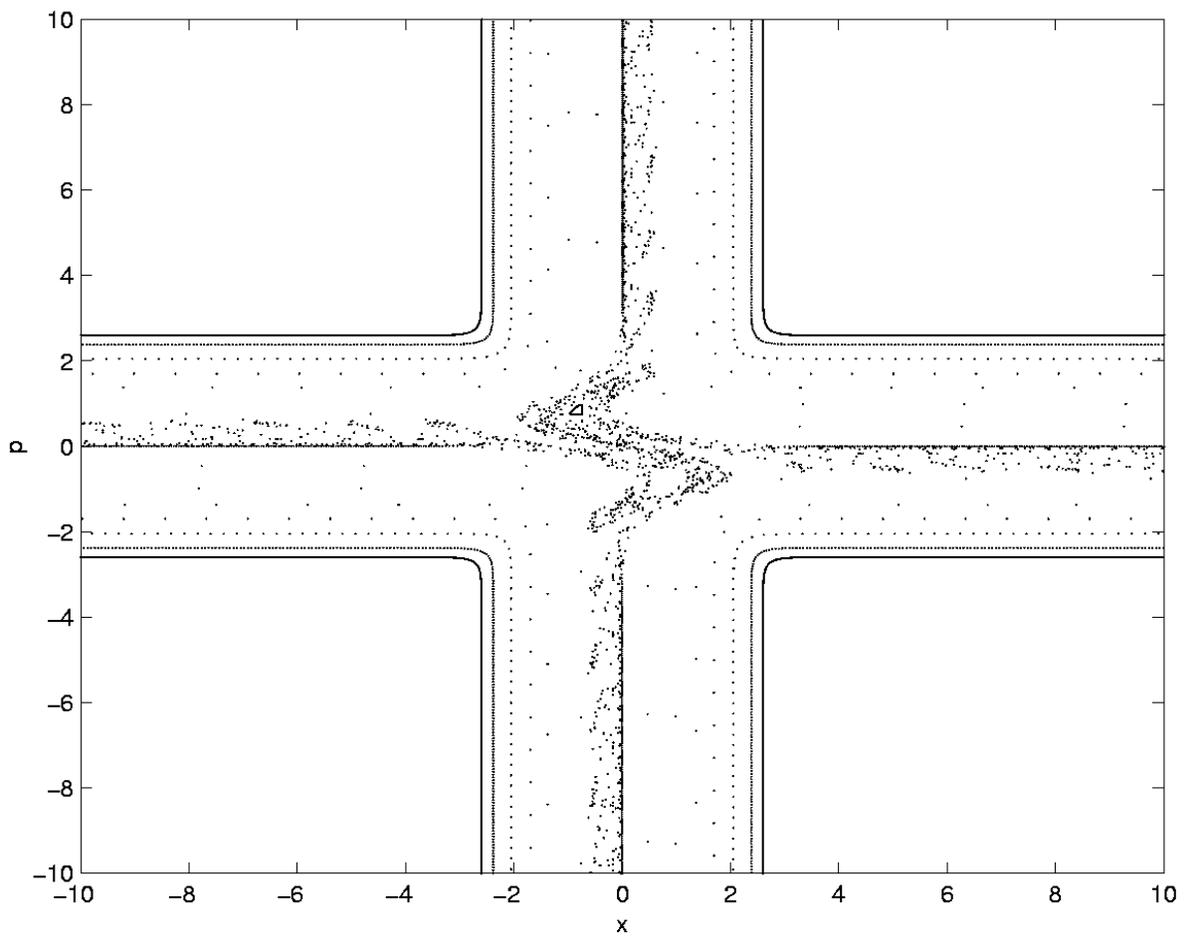
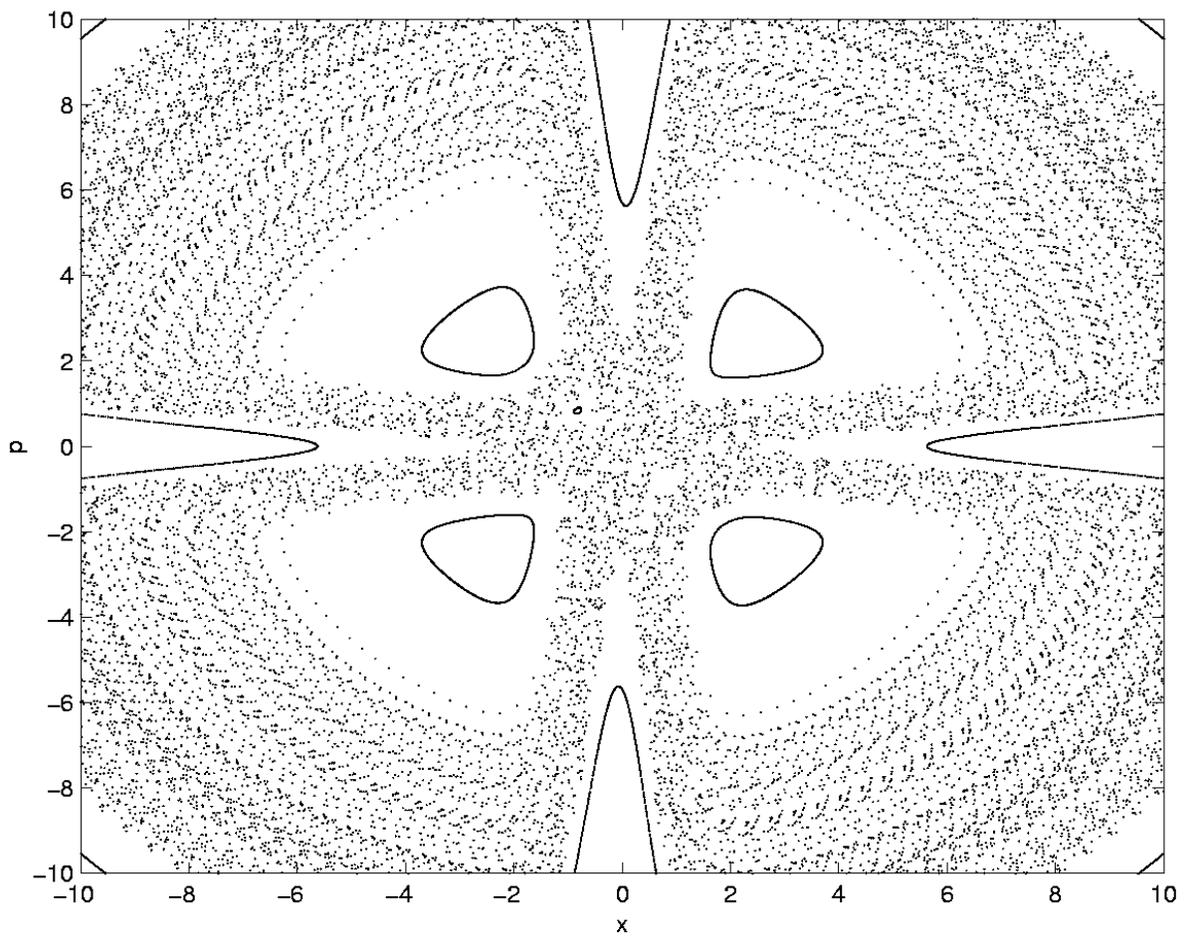
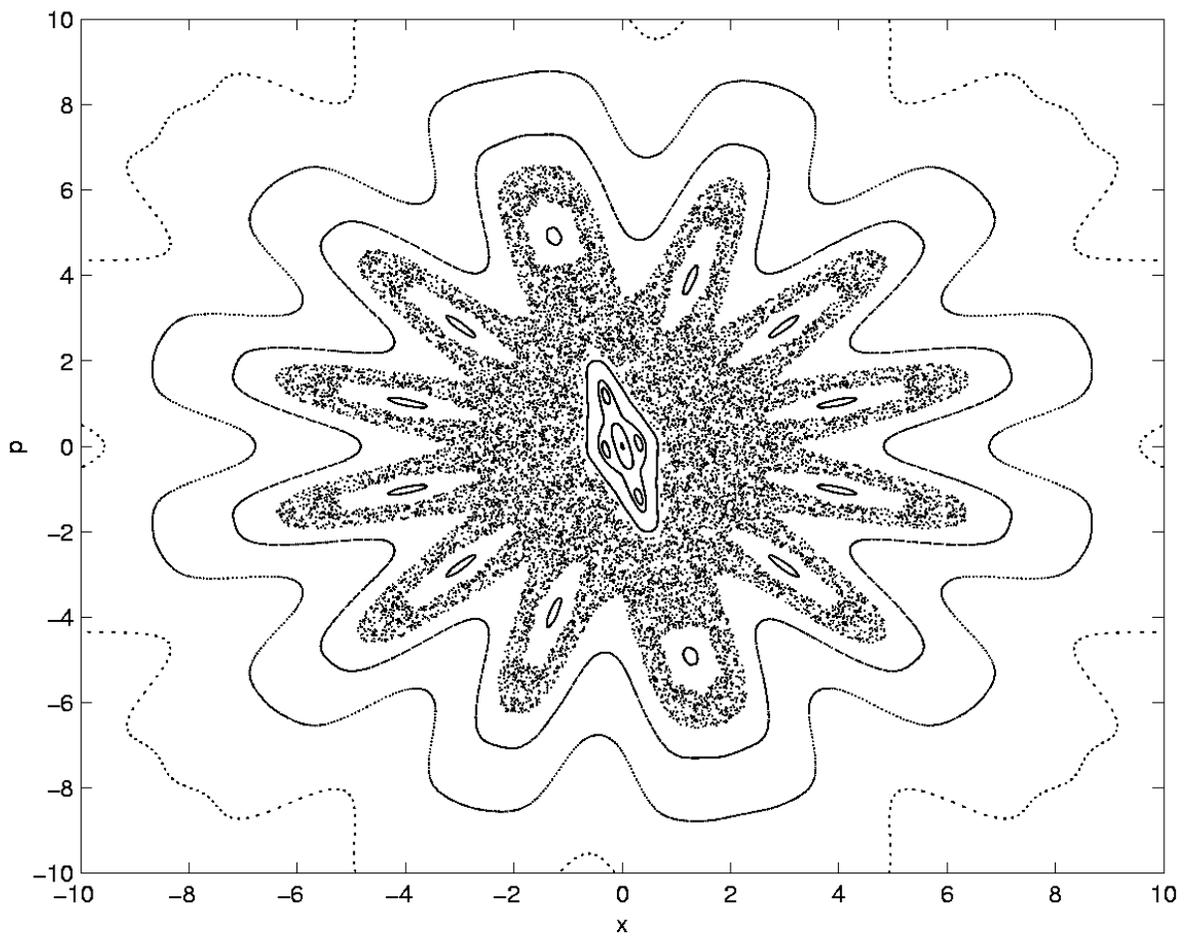
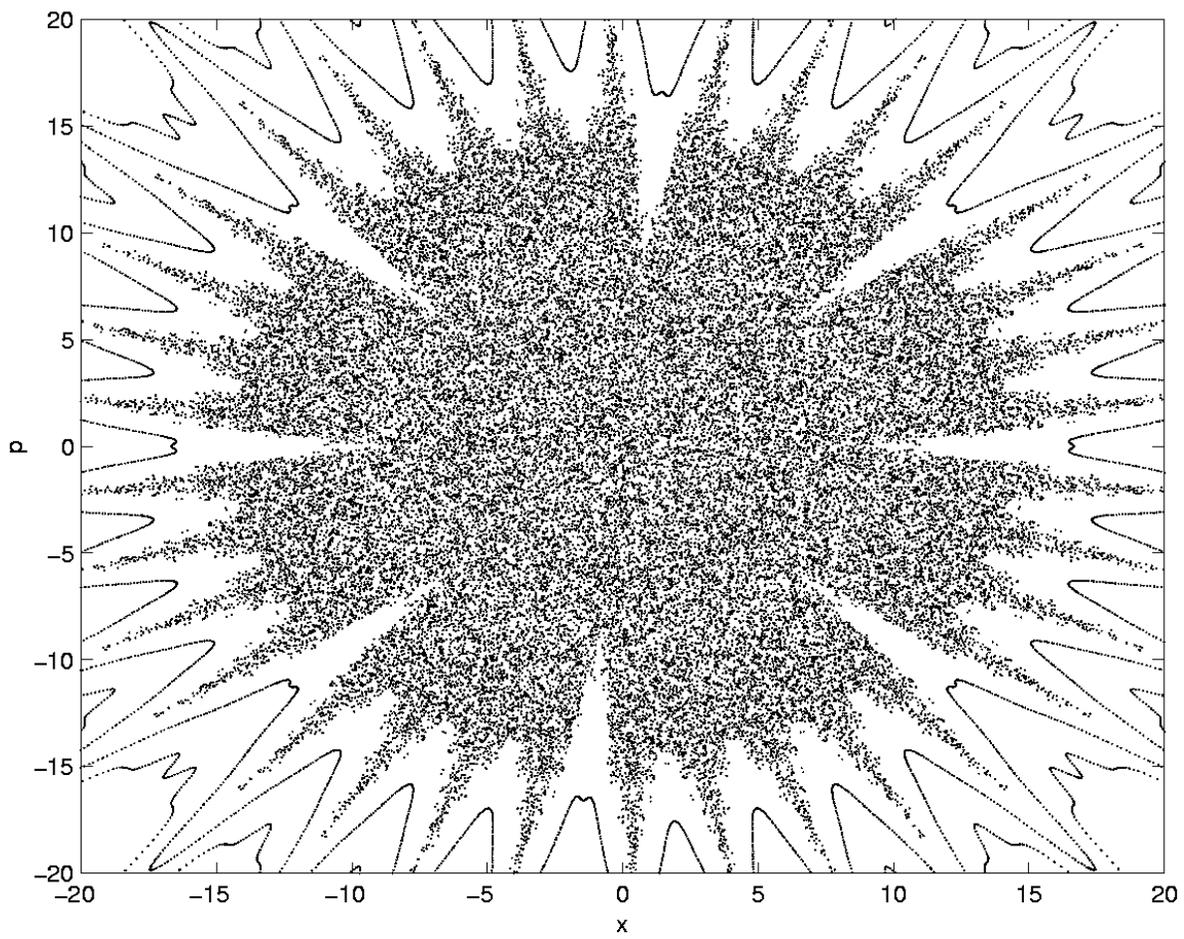
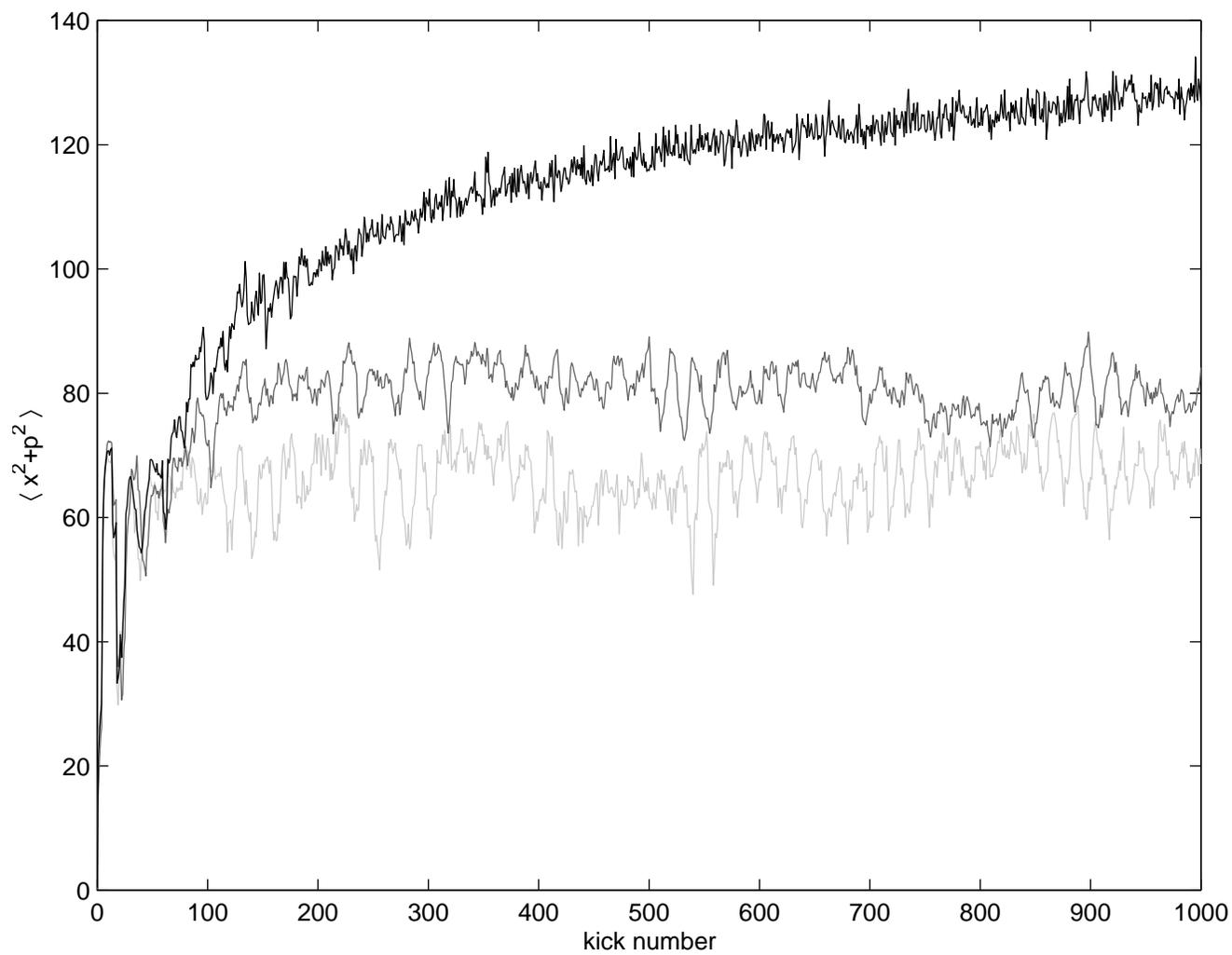
\end{document}